\documentclass[aps,prl,twocolumn,superscriptaddress]{revtex4-2}
\usepackage{graphicx}
\usepackage[export]{adjustbox}
\usepackage[utf8]{inputenc}
\usepackage[T1]{fontenc}
\usepackage{amsmath,amssymb}

\usepackage{xcolor}
\definecolor{prlblue}{RGB}{45, 47, 145}  
\definecolor{prlcite}{RGB}{45, 47, 145} 

\usepackage[colorlinks=true,
linkcolor=prlblue,
citecolor=prlblue,      
urlcolor=prlblue,
breaklinks=true,        
bookmarks=true,
bookmarksnumbered=true,
pdfpagemode=UseNone]{hyperref}  

\begin{document}


\title{Programming Coherent and Quantum Light with a Free-Electron Wavepacket}


\author{Songyu Zhu}
\affiliation{State Key Laboratory of Ultra-intense Laser Science and Technology, Shanghai Institute of Optics and Fine Mechanics (SIOM), Chinese Academy of Sciences (CAS), Shanghai 201800, China}
\affiliation{State Key Laboratory of Quantum Functional Materials, School of Physical Science and Technology and Center for Transformative Science, ShanghaiTech University, Shanghai 200031, China}

\author{Yushan Zeng}
\email{yszeng@siom.ac.cn} 
\affiliation{State Key Laboratory of Ultra-intense Laser Science and Technology, Shanghai Institute of Optics and Fine Mechanics (SIOM), Chinese Academy of Sciences (CAS), Shanghai 201800, China}
\affiliation{Center of Materials Science and Optoelectronics Engineering, University of Chinese Academy of Sciences, Beijing 100049, China}

\author{Chenhao Pan}
\affiliation{State Key Laboratory of Ultra-intense Laser Science and Technology, Shanghai Institute of Optics and Fine Mechanics (SIOM), Chinese Academy of Sciences (CAS), Shanghai 201800, China}
\affiliation{State Key Laboratory of Quantum Functional Materials, School of Physical Science and Technology and Center for Transformative Science, ShanghaiTech University, Shanghai 200031, China}

\author{Yiming Pan}
\email{yiming.pan@shanghaitech.edu.cn} 
\affiliation{State Key Laboratory of Quantum Functional Materials, School of Physical Science and Technology and Center for Transformative Science, ShanghaiTech University, Shanghai 200031, China}

\author{Ye Tian}
\email{tianye@siom.ac.cn} 
\affiliation{State Key Laboratory of Ultra-intense Laser Science and Technology, Shanghai Institute of Optics and Fine Mechanics (SIOM), Chinese Academy of Sciences (CAS), Shanghai 201800, China}
\affiliation{Center of Materials Science and Optoelectronics Engineering, University of Chinese Academy of Sciences, Beijing 100049, China}

\author{Ruxin Li}
\email{ruxinli@siom.ac.cn} 
\affiliation{State Key Laboratory of Ultra-intense Laser Science and Technology, Shanghai Institute of Optics and Fine Mechanics (SIOM), Chinese Academy of Sciences (CAS), Shanghai 201800, China}
\affiliation{State Key Laboratory of Quantum Functional Materials, School of Physical Science and Technology and Center for Transformative Science, ShanghaiTech University, Shanghai 200031, China}
\affiliation{Center of Materials Science and Optoelectronics Engineering, University of Chinese Academy of Sciences, Beijing 100049, China}
\affiliation{Zhangjiang Laboratory, Shanghai 201210, China}


\date{\today}

\begin{abstract}
The pursuit of compact, programmable light sources with high coherence and spectral purity hinges on establishing a precise set of phase relationships in light-matter interactions. Here, we demonstrate that the quadratic dispersion of freely propagating electron wavepacket serves as a programmable quantum medium. Prepared in a coherent momentum-state ladder via a single laser interaction, the electron subsequently undergoes deterministic phase evolution during free propagation—an intrinsic process that compiles its quantum state into two distinct emission channels. This mechanism, quantified by a quantum bunching factor, enables: (i) Talbot-resonant bunching, where the electron density self-structures into sub-cycle combs with tunable harmonic selectivity, and (ii) coherent phase transfer of the programmed quadratic phase to light, generating nonclassical photon states such as multi-component Schrödinger cat states via measurement-conditioned interaction. This quadratic-phase programming establishes a versatile platform for on-demand quantum state synthesis, bridging beam engineering with electron wavefunction shaping for compact quantum light sources, coherent radiation control, and scalable quantum information processing.
\end{abstract}


\maketitle

The quest for intense, coherent radiation and complex photonic quantum states drives modern ultrafast science and quantum technology. This pursuit has spurred diverse physical mechanisms, from high-harmonic generation (HHG)\cite{RN1,RN2,RN3,RN4} and free-electron lasers (FELs)\cite{RN5,RN6} to their envisioned quantum counterparts\cite{RN7,RN8,RN9}. Underpinning these approaches is a common physical requirement: the establishment of a well-defined periodic coherence, whether manifested as electron density modulation or as structured phase-space coherence in the photonic field. 

\begin{figure*}[t] 
	\centering
	\includegraphics[width=1.5\columnwidth,keepaspectratio]{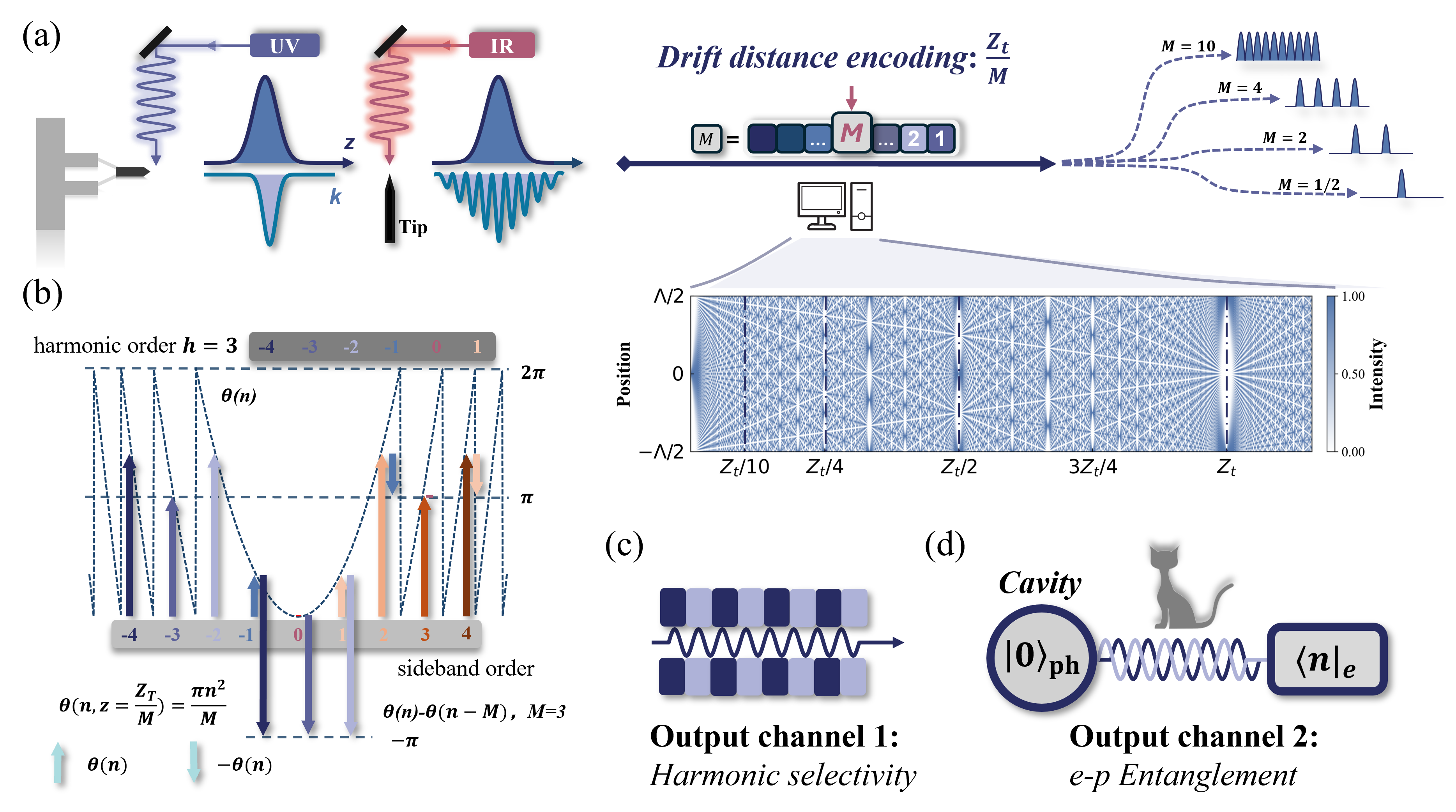}
	\caption{The programmable free-electron wavepacket light source. (a) An electron wavepacket interacts with an optical near-field via PINEM, forming a coherent momentum ladder that undergoes quadratic-phase evolution during free propagation, manifesting the Talbot self-imaging effect (inset). (b) At drift distance $Z_T /M$, the accumulated phase acts as a programmable spectral filter, constructively interfering only with Talbot-resonant harmonics separated by order. (c) Emission channel 1: The compiled electron wavefunction forms a high-contrast density comb, generating programmable high-harmonic attosecond pulse trains. (d) Emission channel 2: The quadratic phase is coherently transferred to an optical mode via measurement, deterministically generating multi-component Schrödinger cat states.}
	\label{1}
\end{figure*}

In advanced seeded FELs, including high-gain harmonic generation (HGHG)\cite{RN10} and echo-enabled harmonic generation (EEHG)\cite{RN11,RN12}, such modulation is imposed externally through cascaded laser modulators and magnetic chicanes. While powerful, these approaches couple modulation depth to spectral bandwidth and achieve high harmonics through multi-stage phase-space manipulation, incurring increased complexity and susceptibility to collective instabilities. A fundamentally different route has been envisioned in the quantum FEL, where spectral purity arises from quantum-recoil-dominated gain, but this demands an exceptionally narrow initial electron energy spread\cite{RN8,RN13}, trading operational robustness for coherence.  This raises a central question: can high-purity, programmable structuring arise intrinsically from the coherent evolution of a quantum wavepacket itself, without external modulators or stringent state preparation?

Here, we explore such a programmable light source by harnessing a ubiquitous yet underutilized resource: the quadratic dispersion of a free electron wavepacket. This mechanism connects to a fundamental wave phenomenon of the wave self-imaging, where fine microstructure self-organizes through quadratic dispersions— exemplified by the Talbot effect\cite{RN14}. Recently,  the Talbot effect has been demonstrated for electron wavepackets\cite{RN15,RN16,RN17}, primarily for generating attosecond pulse trains\cite{RN18,RN19,RN20}. In this letter, we leverage this underutilized potential by using the Talbot effect as a built-in compiler for programmable quantum state synthesis.

As shown in Fig.~1(a), the electron is first prepared in a coherent momentum-state ladder via the interaction of photon-induced near-field electron microscopy (PINEM) \cite{RN18,RN20,RN21,RN22,RN23,RN24,RN25}. Its subsequent free propagation, governed solely by quadratic dispersion, programs this initial state into two distinct emission outputs. First, the electron density self-structures into attosecond pulse trains\cite{RN18,RN22} with Talbot-resonant harmonic selectivity, enabling high-harmonic emission without external seeding or filtering. This constitutes a quantum analogue to beam microbunching, yet emerges from intra-wavepacket interference rather than collective amplification. Second, coherent phase transfer to a quantum cavity mode, conditioned on post-selection, deterministically generates multi-component optical Schrödinger cat states\cite{RN26,RN27}. Both the harmonic order of the attosecond emission and the number of cat-state components are controlled by a single classical parameter: the free-propagation distance. This free-electron wavepacket platform establishes a unified, compact route to programmable high-harmonic and quantum light sources.

\textit{Quadratic-Phase Programming}---The quadratic dispersion of a free electron wavepacket provides the essential resource for this programmable platform. When prepared in a coherent superposition of discrete momentum states, the electron accumulates a deterministic quadratic phase during free propagation—a process that functions as a built-in compiler, controlled solely by the drift distance z. Instead of relying on external modulators or material nonlinearities, our approach leverages this intrinsic yet underutilized resource to sculpt both the density profile of the electron and the quantum state of an interacting radiation field—an approach we term \textit{quadratic-phase programming}.

The programming protocol begins by encoding a coherent momentum-state ladder onto an electron wavepacket. As shown in Fig.~1(a), this can be realized through the PINEM interaction\cite{RN18,RN21}, which prepares a Floquet-Bloch lattice in momentum space: $\psi(z,t=0)=\sum_{l=-N}^Nc_le^{i(p_0+l\hbar k)z/\hbar}$, where $p_0$  is the central momentum, $k=\omega/v_0$ is the photon wave number, and $N$ sets the truncation of the sideband ladder. During subsequent free evolution over a distance $z=v_0 t$, this initial momentum ladder is deterministically “compiled” by the quadratic dispersion governed by the free-space Schrödinger equation, yielding a kinetic phase $\phi_l =[E_0+v_0k\hbar l+(l\hbar k)^2/2m_e\gamma^3]\times z/\hbar v_0$ (see Supplementary Material\cite{SM} for the derivation). The core programming engine is the relative phase accumulated between momentum sidebands  $l$ and  $l+h$ (separated by $h$ quanta):

\begin{equation}
	\begin{split}
		\Delta\phi_{l,h}(z) &\approx \biggl[ v_0hk + \frac{-(l-h)^2\hbar k^2}{2m_e\gamma^3}  + \frac{l^2\hbar k^2}{2m_e\gamma^3} \biggr] \times \frac{z}{v_0} \\
		&= \left[ \frac{2lh\hbar k^{2}}{2m_{e}\gamma^{3}} + \frac{h^{2}\hbar k^{2}}{2m_{e}\gamma^{3}} + v_{0}hk \right] \times \frac{z}{v_{0}} \\
		&= \frac{2\pi lhz}{Z_{T}} + \phi_{h}
	\end{split}
\end{equation}
where $Z_T=2\pi m_\mathrm{e}v^3\gamma^3/\hbar\omega^2$ is the Talbot revival distance, which sets the fundamental scaling of the system; $\phi_h$  is an $l$-independent offset, $m_e$ is the electron mass, and $\gamma$ is the Lorentz factor. Crucially, this phase term $\Delta\phi_{l,h}$ originates purely from free-electron dispersion and therefore functions as a built-in, dynamical phase mask controlled solely by $z$.

To quantify this programming, we employ the (single-particle) bunching factor $b_f (h)$. In classical beam physics, this factor characterizes the amplitude of the $h$-th spatial Fourier harmonic of the ensemble electron density modulation \cite{RN10,RN11}. In the quantum description, it extends naturally to a free-electron wavepacket and captures the periodic order in its probabilistic density: $b_f(h)=|\langle\psi\mid e^{ihkz}\mid\psi\rangle|$, with $\psi$ denoting the electron wavefunction\cite{RN28,RN29}. This bunching factor directly captures the degree of spatiotemporal coherence that a quantum electron wavepacket can impart onto radiation at the corresponding wavelength. 

This quantum bunching, together with the programmed phase in Eq.~(1), reveals the essence of the compilation mechanism:
\begin{equation}
	b_f(h,z)=\left|\sum_lc_l^*c_{l+h}e^{\frac{i2\pi lhz}{Z_T}}\right|
	\label{eq:bf}
\end{equation}
where $c_l$ denotes the amplitude of the $l-th$ momentum sideband. Explicitly, Eq.~(2) coherently sums the contributions from all sideband pairs $(l,l+h)$, showing that the phase term $\phi_{l,h}=2\pi lhz/Z_T$ act as an intrinsic active phase mask programmed solely by $z$. At fractional Talbot distances $z=Z_T/M$, Eq.~(2) becomes resonant: for orders h that are integer multiples of $M$ ($h=lM$ with $l\in\mathbb{Z}$), all contributing terms rephase constructively, yielding a pronounced peak in $b_f$, whereas off-resonant orders are suppressed by destructive interference. Consequently, this self-organized interference of a free-electron wavepacket transforms the classical strategy of engineering collective spatial microbunching in many-electron beams into the quantum challenge of preparing and coherently evolving a free-electron wavepacket in momentum space. As shown below, this quadratic-phase programming establishes $Z_T$ and $z$ as the natural tuning knob of a built-in quantum compiler. It enables two emission channels: tailored electron bunching with intrinsic spectral filtering (Channel 1, Fig.~1(c)), and nonclassical photonic states through coherent interaction (Channel 2, Fig.~1(d)).

\textit{Talbot-Enhanced High Harmonic Bunching}---Building on this programmable scheme, we first demonstrate its capability by sculpting the electron’s own quantum state. This channel directly manifests the Talbot effect as a built-in compiler: the programmed output is a high-contrast, harmonic-selective attosecond electron density modulation—a quantum analogue to the statistically microbunched beams in advanced free-electron lasers. Yet, unlike multi-stage classical schemes that require external filtering, this \textit{Talbot-Enhanced High-Harmonic Bunching} (TEHHB) arises intrinsically from the intra-wavepacket interference of a programmed quantum state. As illustrated in Fig.~1(b), TEHHB directly follows from the resonance structure of the bunching factor in Eq.~(2). At fractional Talbot distances $z=Z_T/M$\cite{RN30}, only harmonic orders satisfying the Talbot condition $h=lM$ rephase constructively, while off-resonant harmonics are suppressed by destructive interference. This intrinsic intra-wavepacket interference establishes a direct and programmable link between the electron’s real-space focusing and its reciprocal-space harmonic spectrum.

\begin{figure}[tb] 
	\centering
	\includegraphics[width=0.9\columnwidth,keepaspectratio]{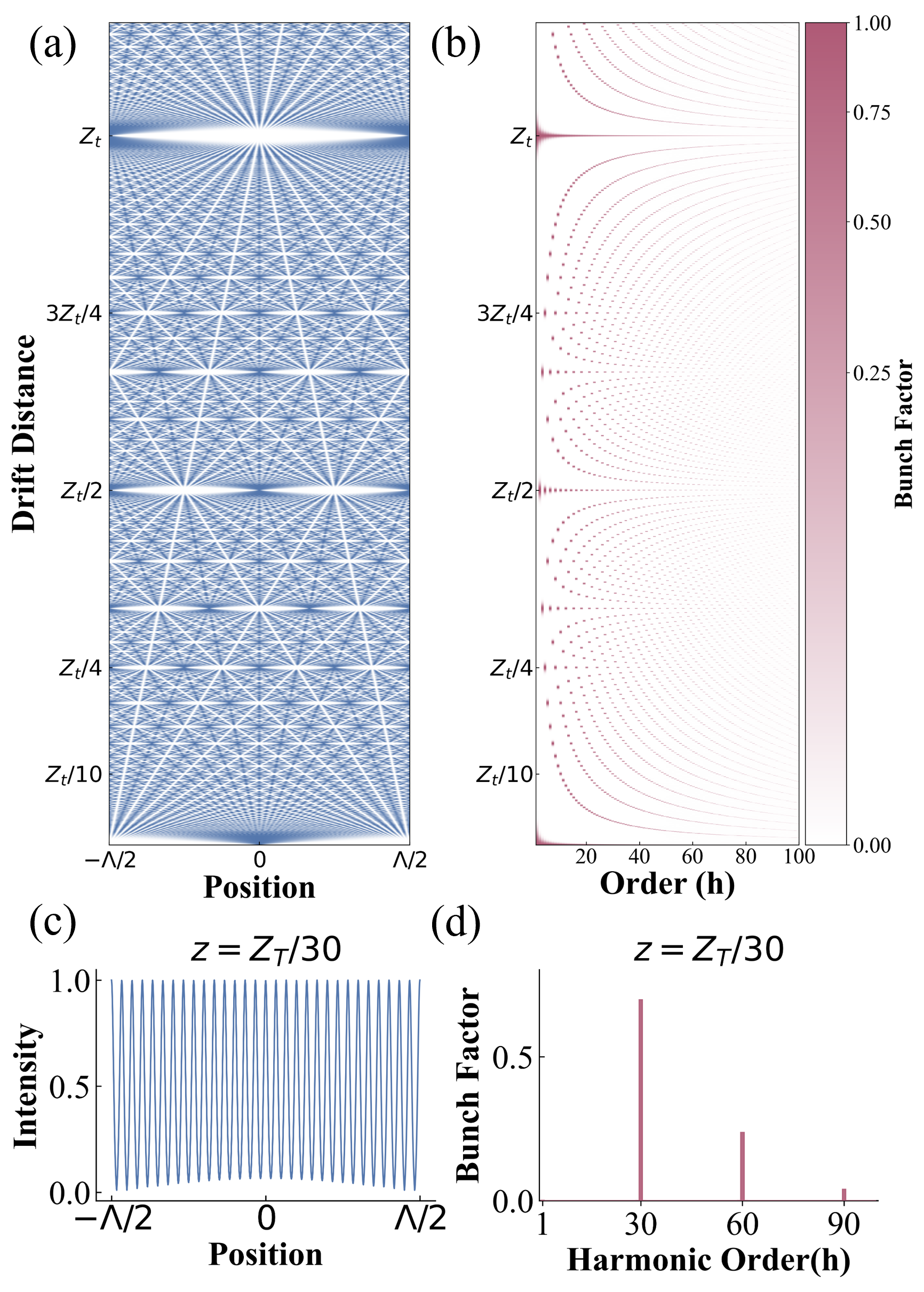}
	\caption{Intrinsic Talbot self-imaging and harmonic bunching of a free-electron wavepacket. (a) Spatiotemporal evolution of electron density showing periodic revivals. Fractional Talbot distances ($z=Z_T/M$) yield subcycle foci; full revival occurs at with phase shift, perfect at $z=Z_T$. (b) Bunching factor evolution with period $Z_T$. (c, d) Density distribution and bunching spectrum at $z=Z_T/30$, demonstrating subcycle localization and selective enhancement of harmonics spaced by order $30$.}
	\label{2}
\end{figure}

Figure 2 shows the spatiotemporal evolution of the programmed electron wavepacket, calculated using experimentally relevant parameters: an electron energy $E_e=200keV$ and a photon-sideband spacing of $1.55 eV$, corresponding to a Talbot distance of $Z_T\approx 238.8mm$. The initial state comprises approximately $N\approx 150$ populated sidebands\cite{RN23}, with $|c_l |^2$ following a Gaussian envelope\cite{RN26}. The free propagation is sampled over the range $[0,1.12^*Z_T]$. As shown in Fig.~2(a), the electron probability density exhibits the characteristic periodic revivals associated with the Talbot effect, and its evolving periodicity is captured by $b_f (h,z)$ (Fig.~2(b)). At fractional distances $z=Z_T/M$, the density profiles partitions into $M$ sharply localized attosecond peaks per optical cycle, forming the interference pattern known as a “quantum carpet”\cite{RN30,RN31}, as shown in the representative frames of Fig.~2(c). Correspondingly, the bunching spectrum (Fig.~2(d)) becomes highly selective, exhibiting peaks only at harmonic orders that are integer multiples of $M$, while off-resonant orders are supressed—consistent with the Talbot phase matching condition in Eqs.~(1) and (2).

The peak width $\Delta x\sim\lambda/N$ reaches the attosecond scale for these parameters and is ultimately bounded by the Fourier constraints set by the populated sideband bandwidth  $\Delta p\sim N\hbar k$. It demonstrates that the coherent electron wavepacket can deterministically self-organize into a periodic train of attosecond bunches emerging purely from intrinsic single-particle quantum interference rather than collective amplification.

The origin of this sharp spectral selection is illustrated in Fig.~3, which examines the momentum-space correlation $\psi^*(p)\psi(p+\hbar hk)$ that constitutes the summation in Eq.~(2). For an off-resonant harmonic (e.g., $h=2$ at $z=Z_T/30$), the programmed phase varies rapidly with sideband index $l$, so contributions from different sideband pairs accumulate mismatched phases and interference destructively (Figs.~3(a), (b)).  In contrast, for the resonant harmonic ($h=30$), the phase is uniform across all dominant contributions, with $exp(i2\pi l)\equiv 1$, yielding coherent alignment and a pronounced peak in  $b_f$ (Figs.~3(c), (d)). This comparison shows that harmonic selectivity emerges precisely when sideband pairs satisfy the Talbot phase-matching condition, directly linking the observed spectral purity to the deterministic quantum interference (Fig.~2).

This mechanism represents a fundamental departure from classical EEHG/HGHG schemes, which rely on collective phase-space manipulation to generate a dense harmonic spectrum and enforce selectivity only at the radiation stage through finite FEL gain bandwidth. In such classical schemes, spectral purity emerges from gain competition among multiple coexisting harmonics during amplification. In contrast, TEHHB embeds selectivity at the quantum state preparation stage: the engineered momentum comb undergoes deterministic Talbot evolution that coherently rephases a discrete set of harmonics while driving off-resonant orders into destructive interference prior to any radiation or gain narrowing. This quantum interference-based selection occurs intrinsically within the single-electron wavepacket, providing a deterministic route to spectrally pure emission. We show in the End Matter that, this selectivity proves robust against experimental imperfections: stochastic phase noise raises the off-resonant noise floor without erasing resonant peaks, and finite sideband energy spread remains within the coherence length at fractional Talbot distances where selective bunching is optimized (e.g., $z=Z_T/30$), preserving the programmed interference pattern.

\begin{figure}[tb] 
	\centering
	\includegraphics[width=\columnwidth,keepaspectratio]{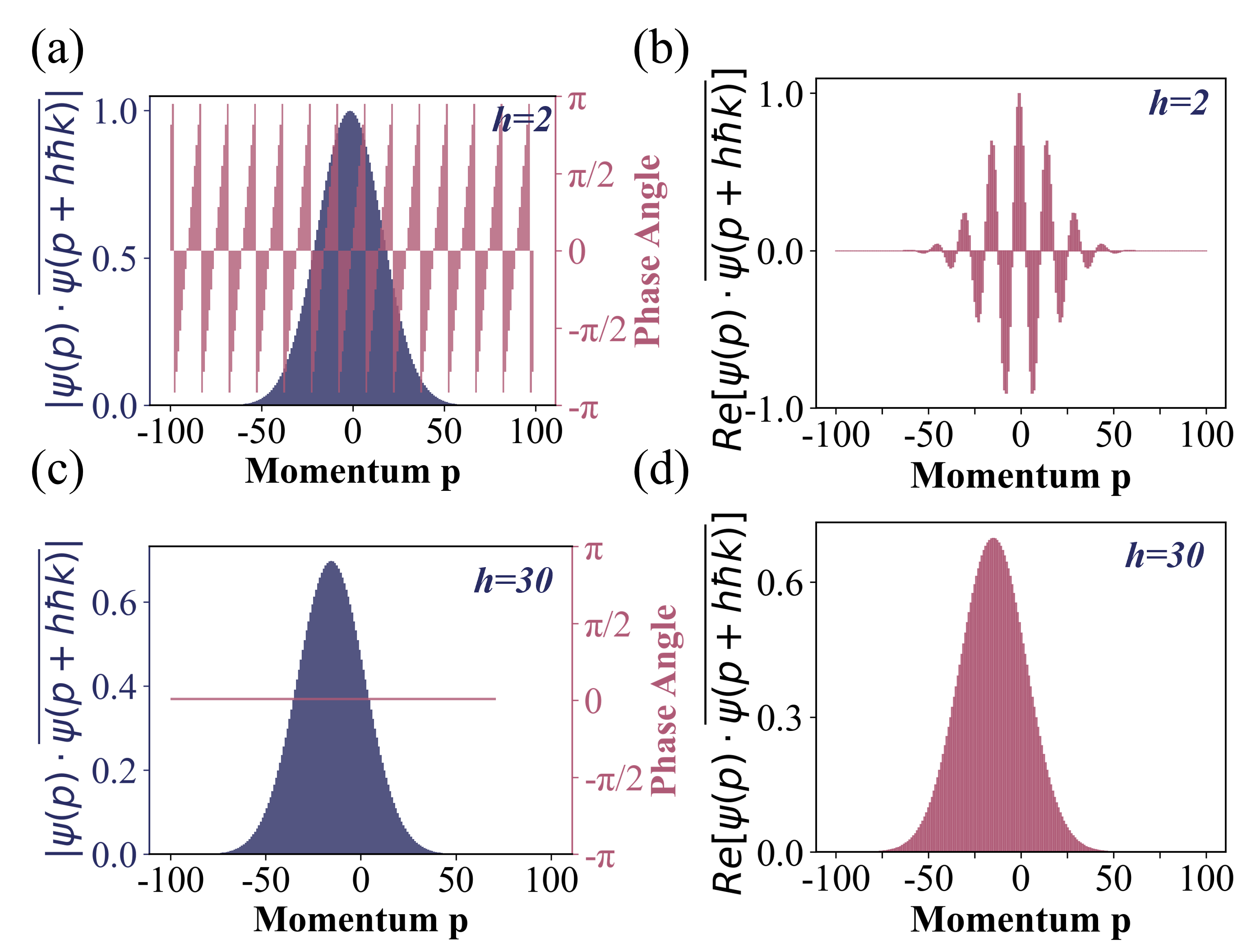}
	\caption{Momentum-space origin of selective harmonic bunching. (a-b) At off-resonant order $h=2$, phase oscillations among momentum sidebands cause destructive interference, suppressing bunching. (c-d) At resonant order $h=30$, phase alignment ($\Delta \phi \approx 0$) enables constructive interference and pronounced enhancement. This phase-matching condition underpins the order-selective dynamics observed in Fig.~2.}
	\label{3}
\end{figure}

\textit{Synthesizing Optical Cat States}---Quadratic-phase programming is not restricted to structuring the electron wavepacket itself; rather, it represents a coherent phase resource that can be transferred to external degrees of freedom. Crucially, the phase mask $\phi_{l,h}=2\pi lhz/Z_T$ can be coherently transferred from the momentum-space lattice onto a radiation field via interaction and measurement. This idea casts the electron not only as a compiler of its own probability density, but also as a universal phase mediator that extends the programming paradigm from wavepacket self-imaging to nonclassical photon state synthesis.

We now examine the second emission channel of this programmable scheme: using the modulated electron to synthesize nonclassical states of light. We couple the electron to a quantized optical mode (e.g., a high-Q cavity initially in vacuum) and performing a post-selective measurement, which conditionally transfers the programmed quadratic phase to the photon statistics of light field. The scheme is illustrated in Fig.~1(d). After the coherent electron-photon interaction, the joint state becomes entangled \cite{RN26,RN29,RN32}, a subsequent post-selection of the electron energy projects the joint state onto a conditional photon state. Specifically, when the electron is projected in the momentum (energy) sideband indexed by $n^{\prime}$, the resulting photon state reads 
\begin{equation}
|\phi_{n^{\prime}}\rangle=\sum_{m=0}^\infty e^{i\pi\frac{(n^{\prime}+m)^2}{M}}e^{-\frac{|g|^2}{2}}\frac{g^m}{\sqrt{m!}}|m\rangle_{ph}
\label{eq:phase}
\end{equation}
where $m$ is the photon number in the Fock basis $|m\rangle_{\mathrm{ph}}$, $g$ is the dimensionless electro–photon coupling strength. The programming factor $e^{i\pi(n^{\prime}+m)^{2}/M}$ emerges directly from the accumulated phase $\mathrm{e}^{\mathrm{i}\pi n^2/M}$ with the quadratic phase structure initially defined over the electronic sideband index being conditionally inherited by the photon number basis through the interaction and projection. Importantly, the imprinting does not depend on the particular detected electron state, ensuring that the programmed phase information is universally encoded onto the conditional photon state.

The conditional state $\mid\phi_{n^{\prime}}\rangle$ constitutes a coherent superposition of multiple phase-rotated components in optical phase space, i.e., a programmable multi-component Schrödinger cat state. The programming parameter M, set by the drift distance $z =Z_T/M$, determines the number of distinct coherent state components in the superposition. Figure 4 presents the Wigner quasiprobability distributions for synthesized states with varying $M$. The phase-space patterns unambiguously exhibit nonclassical interference fringes and discrete nodal structures characteristic of cat states, with the number of major lobes scaling directly with $M$.

This result demonstrates a paradigm shift from material (Kerr)\cite{RN33,RN34} to geometric (dispersion-induced) quantum nonlinearity. Unlike conventional methods that rely on the weak and lossy optical Kerr effect in optical media, our approach harnesses the deterministic, lossless quadratic phase accumulated by a free electron in vacuum as the effective nonlinear engine. The programming is exceptionally versatile: by tuning a single classical parameter of drift distance $z$, the topology of the quantum light state can be reconfigured, changing both the number of cat-state components and their phase-space arrangement in real time. This control is directly analogous to the harmonic selectivity in TEHHB, where the same parameter $z$ selects the enhanced harmonic order $h=lM$. Here, it selects the superposition multiplicity $M$. 


\begin{figure}[tb] 
	\centering
	\includegraphics[width=0.8\columnwidth,keepaspectratio]{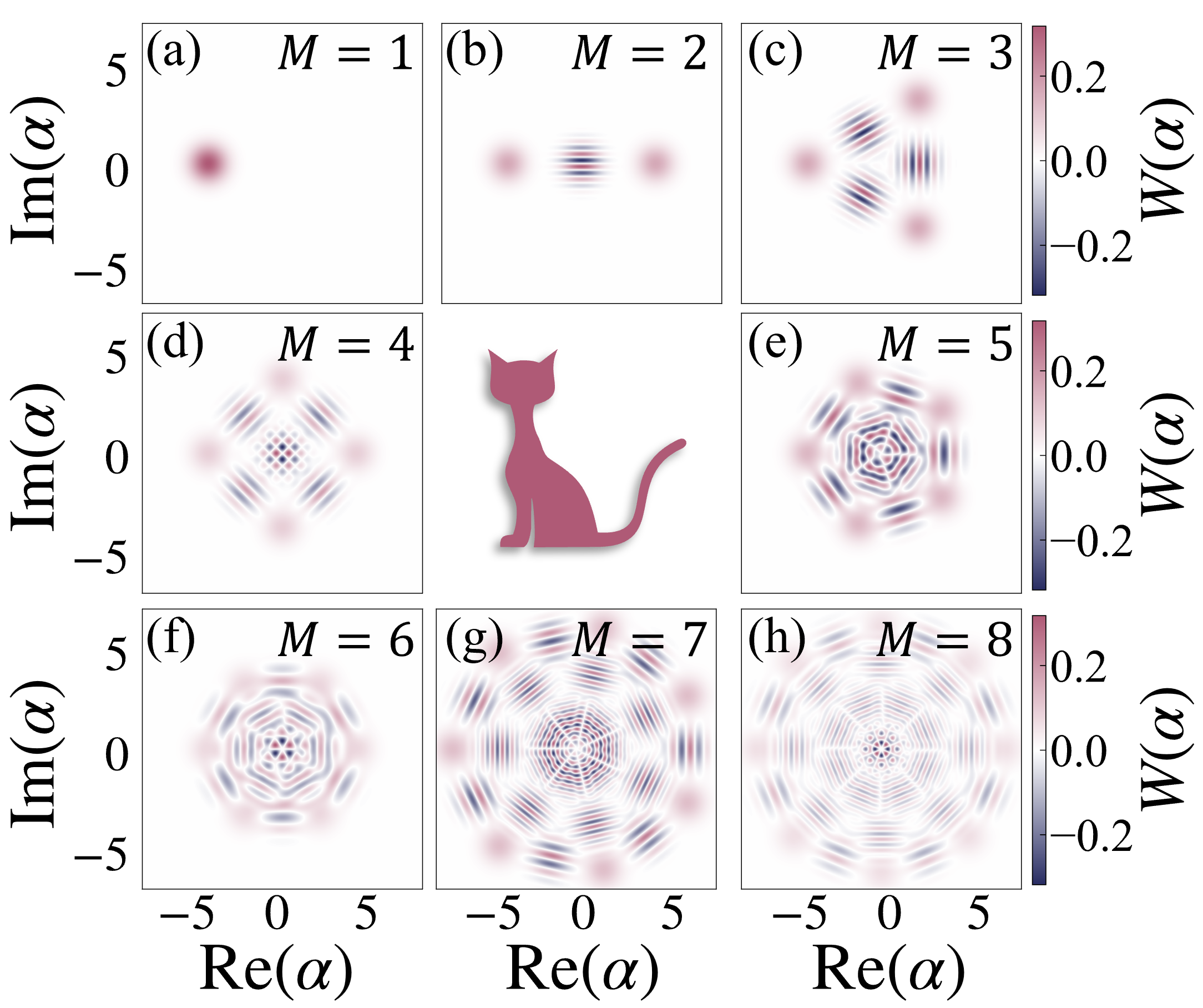}
	\caption{Preparation of Schrödinger-cat states mediated by Talbot-modulated free electrons. (a–f) Wigner quasiprobability distributions of the generated photonic states for Talbot indices ranging from $M=1$ to $M=6$, respectively, with the coupling strength fixed at $g=3$. The phase-space structures clearly reveal coherent superpositions of M distinct coherent components, including Schrödinger-cat–like states. (g,h) Same as (a–f), but with an increased coupling strength $g=5$.}
	\label{4}
\end{figure}


\textit{Conclusion}---We demonstrate a programmable free-electron wavepacket platform that emits both high-harmonic and quantum light through intrinsic dispersive evolution—a paradigm shift from material (Kerr) to geometric (dispersion-induced) nonlinearity. By embedding spectral selectivity into quadratic-phase programming, this approach harnesses deterministic, lossless phase accumulation as the effective nonlinear engine, transforming external filtering into self-compiled, narrow-band electron modulation. Operating as a dual-channel system controlled by drift distance $z$ , the programmed phase kernel $\exp{(i2\pi lhz/Z_T)}$ yields distinct outputs: when measured in real space, it acts as an interference filter generating an attosecond electron comb (Channel 1); when projected onto momentum sidebands, the phase coherently transfers to photon Fock states, producing programmable M -component Schrödinger cat states (Channel 2). This unified framework establishes a template for quantum state synthesis applicable to any quadratic-dispersion system. For free electrons specifically, it offers compact coherent short-wavelength sources and on-chip quantum light generation\cite{RN35,RN36}, including squeezed light and cat states, with spectral selectivity encoded in wavefunction evolution rather than external resonators. By treating free electrons as quantum registers rather than mere radiation sources, this scheme opens pathways toward hybrid quantum information processing\cite{RN17,RN37} architectures that unify computation, coherent light generation, and ultrafast spectroscopy.

\textit{Acknowledgments}---T. Y. acknowledges the support of the National Natural Science Foundation of China (12388102, 12325409) and Shanghai Pilot Program for Basic Research, Chinese Academy of Sciences, Shanghai Branch. Y. P. acknowledges the support of the NSFC (No. 2023X0201-417-03) and the fund of the ShanghaiTech University (Start-up funding).

\textit{Data availability}---All data in the paper is available from the corresponding author on reasonable request.

\bibliography{bibTex}

\onecolumngrid 
\section*{End Matter}
\twocolumngrid 

\vspace{0.5em} 

In this section, based on the quantum interference framework established above, we construct a robustness analysis of quadratic-phase programming against realistic experimental imperfections. Through quantitative examination of stochastic phase noise and finite sideband energy spread under the experimental conditions used in Figs.~2--3 ($E_e=200keV$, $N\approx 150$ sidebands, $z=Z_T/30$ ), we demonstrate how Talbot-enabled harmonic selectivity emerges as a resilient component within this programmable scheme.

The robustness of this programmable scheme rests on preserving relative phase coherence between sidebands, rather than absolute phase stability. Consequently, a global phase ramp---such as the $n\theta$  term imprinted in a standard PINEM interaction\cite{RN24}---induces only a spatial translation of the entire electron wavepacket, leaving all inter-sideband phase relationships intact (see Supplementary Material\cite{SM}).

Stochastic phase noise, however, directly randomizes the relative phase $arg(c_n)$ , perturbing the precise destructive interference designed to suppress off-resonant harmonics. As shown in Fig.~5(a), this raises a noise floor across the bunching spectrum while the resonantly enhanced harmonics (e.g.,$h=30$ at $z=Z_T/30$) remain dominant. The signal-to-noise ratio decreases monotonically with increasing disorder but remains substantial even under strong noise (Fig.~5(b)), demonstrating that while stochastic noise smears the sharp destructive interference minima, it does not erase the fundamental harmonic selectivity encoded by the quadratic phase resonance.

Finite sideband width presents a second practical limitation, introducing a distribution of longitudinal velocities that causes phase slippage during propagation. While this progressively degrades Talbot revivals at long distances [compare Fig.~5(c) at $z=2Z_T$], the high-harmonic selectivity is optimized at specific, short fractional Talbot distances (e.g., $z=Z_T/30$ ) where phase dispersion remains well within the coherence length of the prepared momentum-state lattice. As shown in Fig.~5(d), the real-space density profiles for zero and realistic finite sideband width ($FWHM_{sideband}=0.15\hbar\omega$ ) are nearly indistinguishable at these operating points.

These results confirm that while long-range self-imaging may be progressively washed out, the phase coherence responsible for harmonic selection is preserved under standard experimental conditions. The scheme does not require cryogenic isolation or coherence beyond existing ultrafast electron microscopy, ensuring the feasibility of robust, integrable radiation sources based on quadratic-phase programming.

\begin{figure}[tb] 
	\centering
	\includegraphics[width=\columnwidth,keepaspectratio]{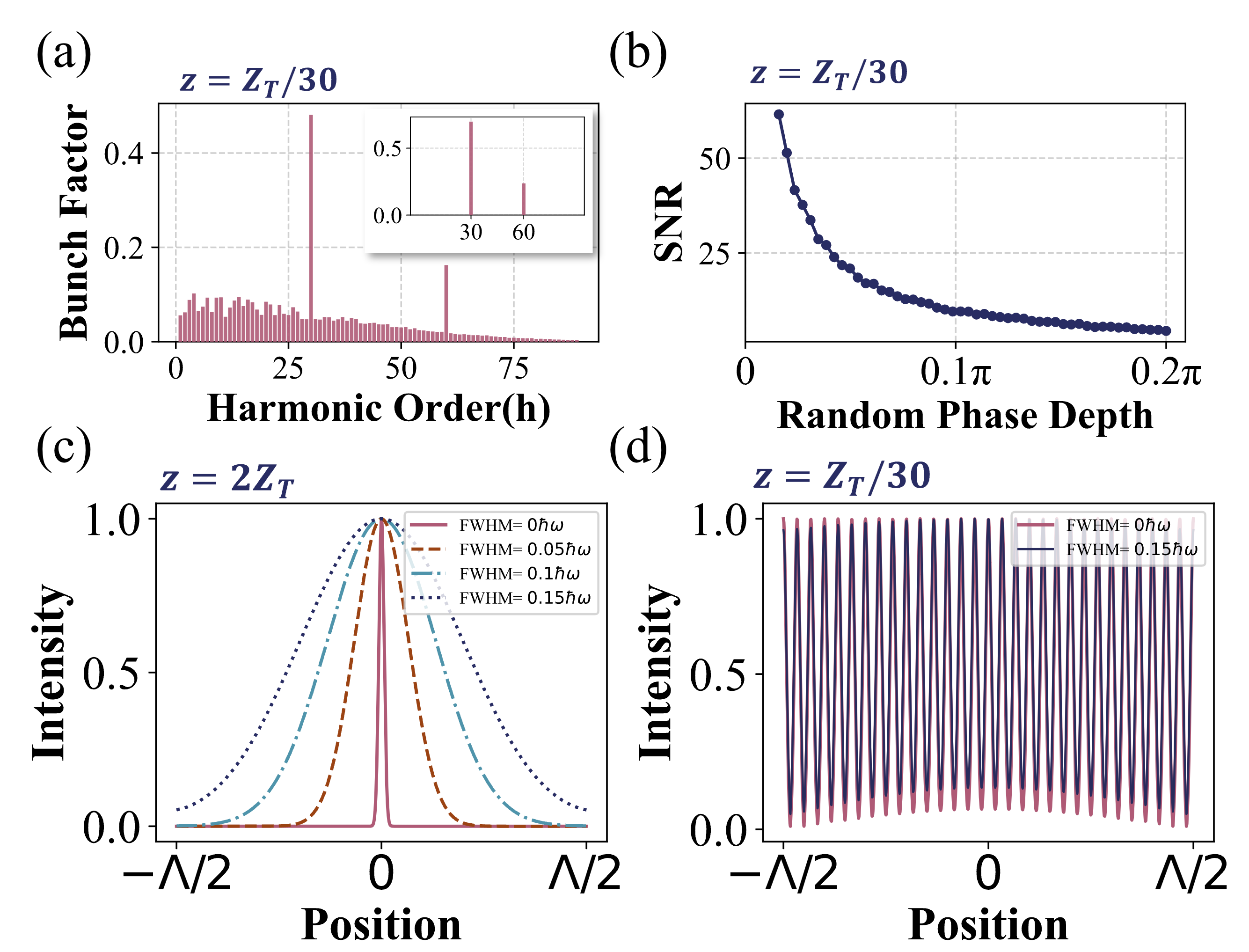}
	\caption{Robustness of harmonic bunching against phase noise and energy spread. (a) Bunching spectrum at $z=Z_T/30$ with phase noise $0.2\pi$ (inset: ideal case). (b) SNR degrades monotonically with increasing phase noise depth. (c) At $z=2Z_T$, sideband widths $FWHM =0,0.05,0.1,0.15\hbar \omega$    progressively degrade real-space reconstruction from near-perfect (narrow) to broadened profiles. (d) At $z=Z_T/30$, distributions for $FWHM = 0\hbar\omega$ and $0.15\hbar\omega$ remain nearly identical, indicating that harmonic-selective bunching is preserved despite energy spread over short propagation. }
	\label{5}
\end{figure}

\end{document}